\documentclass[11pt,a4paper]{article}
\usepackage[utf8]{inputenc}
\usepackage{graphicx}
\usepackage{amsmath,amssymb,mathtools}
\usepackage{hyperref}
\hypersetup{colorlinks=true,linkcolor=blue,citecolor=blue,urlcolor=blue}
\usepackage{geometry}
\usepackage{multirow}
\date{}

\begin{document}

\title{{\bf Classical and quantum perspectives on temperature in accelerated frames}}

\author{Babak Vakili\thanks{email:
ba.vakili@iau.ac.ir}\\\\{\small {\it
Department of Physics, CT.C., Islamic Azad
University, Tehran, Iran}}} \maketitle

\begin{abstract}
We study the notion of temperature in uniformly accelerated frames within a relativistic thermodynamic framework. Considering a perfect fluid at rest in a non-inertial (Rindler) frame, we derive the condition for thermal equilibrium from energy--momentum conservation. This leads to a position-dependent temperature profile consistent with the Tolman--Ehrenfest relation. We emphasize that this temperature characterizes the local thermodynamic equilibrium of the fluid in the accelerated frame and does not arise from a transformation law between different observers. The physical interpretation of the result and its relation to acceleration-induced effects are briefly discussed.
\vspace{5mm}\noindent
\end{abstract}

\section{Introduction}

The transformation properties of temperature in relativistic settings have been a subject of long-standing debate since the early days of relativistic thermodynamics. While special relativity provides a clear framework for the transformation of space and time between inertial observers, the status of temperature under relativistic transformations has remained subtle and, in some contexts, controversial. Early works by Einstein and Planck suggested that temperature transforms under Lorentz boosts, whereas alternative formulations later proposed temperature as a Lorentz scalar or introduced modified transformation laws \cite{einstein1907, planck1908, ott1963, landsberg1966, Req}.

In recent decades, renewed interest in relativistic thermodynamics has emerged, motivated by developments in high-energy physics, relativistic fluids and cosmology \cite{Wal, Syl, Tad}. Several consistent covariant formulations of thermodynamics have been proposed, emphasizing the role of local equilibrium and the energy--momentum tensor of matter \cite{eckart1940,israel1979}. Within this framework, temperature is naturally defined as a local scalar field associated with a fluid element, rather than a global quantity subject to observer-dependent transformation rules.

An important extension of relativistic thermodynamics arises when considering non-inertial observers. In curved spacetimes or accelerated frames, gravitational and inertial effects influence thermal equilibrium. A classic result in this context is the Tolman--Ehrenfest relation, which states that thermal equilibrium in a static gravitational field requires a spatially varying temperature profile \cite{tolman1930,tolman1934}. This result highlights that temperature gradients may arise even in equilibrium situations, provided that the system is subject to acceleration or gravity. A related modern discussion of the thermal properties of accelerated systems was given by Requardt \cite{requardt2013}, who revisited the Tolman--Ehrenfest temperature relation from an entropy-maximum principle and discussed its connection with the thermal interpretation of the Unruh effect. In particular, the spatial dependence of temperature in a static gravitational field was derived in detail from the requirement of maximal entropy at equilibrium. Our approach is complementary to this analysis. Rather than starting from an entropy-maximization principle, we derive the equilibrium temperature profile directly from the covariant conservation of the energy--momentum tensor for a perfect fluid at rest in a uniformly accelerated frame. This provides a hydrodynamic formulation of the equilibrium condition and allows us to examine explicitly the role of the fluid pressure and equation of state.

A particularly simple and instructive setting for studying accelerated observers is provided by the Rindler frame, which describes uniformly accelerated observers in flat Minkowski spacetime. Despite the absence of spacetime curvature, the non-inertial nature of the Rindler frame leads to effects analogous to those encountered in gravitational fields. The Rindler metric therefore offers a natural arena for investigating thermodynamic equilibrium from the perspective of accelerated observers.

In our previous works \cite{Vakili1, Vakili2}, we investigated the relativistic transformation of temperature between inertial observers by considering specific fluid models with well-defined equations of state, such as relativistic gases, photon gases, perfect fluids with negative presure and generalized Chaplygin gases. In those studies, a fluid at rest in one inertial frame was assigned a temperature and the temperature measured by another inertial observer moving with constant velocity was derived using Lorentz transformations and thermodynamic consistency conditions. The focus there was explicitly on observer-to-observer transformation laws for temperature in inertial frames.

The present work adopts a fundamentally different perspective. Rather than comparing temperatures measured by different observers for the same fluid state, we analyze the condition for thermal equilibrium of a perfect fluid that is at rest in a uniformly accelerated frame, see also \cite{Blbod}. The temperature field obtained in this approach is not the result of a transformation law between frames, but instead emerges from the requirement of local thermodynamic equilibrium in the presence of acceleration. In particular, the temperature profile is derived from energy--momentum conservation and the relativistic Euler equation in a non-inertial setting.

Within this framework, the temperature depends on the spatial coordinate associated with the Rindler frame and reflects the influence of acceleration on equilibrium conditions. This approach clarifies that, in accelerated frames, the notion of temperature is inherently local and tied to the fluid's four-velocity field rather than to a global observer-dependent quantity. Although quantum effects such as the Unruh phenomenon provide an important complementary perspective on acceleration and temperature, the present analysis remains primarily classical and thermodynamic in nature.

\section{Thermodynamics in non-inertial frames}

In this section we briefly review the formulation of relativistic thermodynamics in non-inertial (accelerated) frames, emphasizing the notion of local thermal equilibrium and the observer-dependence of temperature. This discussion will provide the conceptual and technical basis for the explicit analysis in accelerated coordinates presented in the next section.

In a general non-inertial frame, or equivalently in a curved spacetime, the notion of global thermodynamic equilibrium is no longer meaningful. Instead, one assumes the existence of \emph{local thermal equilibrium}, whereby thermodynamic quantities are well-defined only locally and vary smoothly over spacetime. In this framework, the entropy current is written as

\begin{equation}
s^{\mu} = s\, u^{\mu},
\end{equation}
where $s$ is the entropy density measured in the local rest frame of the fluid and $u^{\mu}$ is the four-velocity field of the fluid elements. Local equilibrium is characterized by the condition

\begin{equation}
\nabla_{\mu} s^{\mu} = 0,
\end{equation}
which expresses the absence of entropy production in equilibrium configurations. Accordingly, temperature $T(x)$, energy density $\rho(x)$ and pressure $p(x)$ are treated as local quantities defined with respect to the comoving observer. This assumption constitutes the minimal thermodynamic input and does not rely on any specific equation of state.

A key feature of relativistic thermodynamics is that temperature is not an invariant scalar but an observer-dependent quantity. In a local equilibrium state, the one-particle distribution function may be written in the covariant form

\begin{equation}
f(x,p) \propto \exp\!\left[-\beta(x)\, u_{\mu}(x)\, p^{\mu}\right],
\end{equation}
where $\beta(x) = 1/T(x)$ is the inverse local temperature. It is important to distinguish between two conceptually different notions of temperature:
\begin{itemize}
\item[(i)] the \emph{thermodynamic temperature} of the fluid, defined through the local Gibbs distribution and the entropy density;
\item[(ii)] the \emph{operational temperature} measured by a given observer or detector, which may depend explicitly on the observer's state of motion.
\end{itemize}
Even in local thermal equilibrium, these two notions do not necessarily coincide, a point that becomes particularly relevant in accelerated frames.

Let us consider a stationary spacetime with metric

\begin{equation}
ds^2 = -g_{00}(\mathbf{x})\, dt^2 + g_{ij}(\mathbf{x})\, dx^i dx^j ,
\end{equation}
where $g_{00}$ is independent of time. For a fluid in thermal equilibrium with respect to the timelike Killing vector $\partial_t$, the condition of vanishing heat flux leads to the Tolman--Ehrenfest relation \cite{tolman1934}

\begin{equation}
T(\mathbf{x})\, \sqrt{-g_{00}(\mathbf{x})} = \text{const}.
\end{equation}
This relation implies that, in equilibrium, the local temperature is redshifted by the gravitational (or inertial) potential. Importantly, the Tolman law follows solely from the existence of a stationary background and local thermal equilibrium and is therefore independent of the specific equation of state of the fluid.

Since acceleration may be considered as an effective gravitational field, using the equivalence principle, the Tolman--Ehrenfest law admits a natural interpretation in terms of acceleration. Defining the four-acceleration of the stationary observers as

\begin{equation}\label{4-a}
a_{\mu} = u^{\nu} \nabla_{\nu} u_{\mu},
\end{equation}
one finds that, in equilibrium

\begin{equation}\label{A}
\nabla_i \ln T = - a_i .
\end{equation}
Thus, the spatial variation of the temperature precisely compensates the presence of acceleration, ensuring the absence of heat flow in the accelerated frame.
This observation highlights the close analogy between gravitational fields and non-inertial frames and prepares the ground for the explicit analysis of uniformly accelerated observers.

The Tolman--Ehrenfest relation applies to the thermodynamic temperature of matter in local equilibrium and should not be confused with temperatures associated with quantum vacuum effects, such as the Unruh temperature perceived by accelerated detectors \cite{Unru1, Unru2}. Moreover, in the presence of horizons, particle creation or strong time dependence, local thermal equilibrium may break down and the notion of temperature requires further qualification \cite{Unru3}. 

The physical interpretation of the Unruh temperature, however, remains a subject of discussion. Buchholz and Solveen \cite{Unru3} questioned the interpretation of the Unruh effect as indicating a genuine thermal environment surrounding an accelerated observer. In a related analysis, they argued that the temperature of the Minkowski vacuum itself should remain zero, and that the thermal response of an accelerated detector results from its quantum coupling to the vacuum rather than from an exchange of heat with a thermal medium \cite{buchholz2016}. This interpretation has been challenged by Requardt \cite{requardt2013}, who argued in favor of the physical reality of the Unruh temperature and emphasized its compatibility with the thermal properties associated with accelerated frames. We do not attempt to resolve this interpretational issue here. Our purpose is more limited: to distinguish the classical temperature associated with local thermodynamic equilibrium of a perfect fluid in an accelerated frame from the detector-based temperature appearing in the quantum Unruh effect.
In the following section, we specialize the above general considerations to uniformly accelerated frames described by the Rindler metric, where these issues can be analyzed explicitly.

\section{Uniformly accelerated frames and the Rindler metric}

We now specialize the general considerations of the previous section to uniformly accelerated observers in flat spacetime. This provides a concrete realization of non-inertial frames and allows for an explicit derivation of the Tolman--Ehrenfest law in an accelerated setting.

Consider Minkowski spacetime with line element

\begin{equation}
ds^2 = -dT^2 + dX^2 + dY^2 + dZ^2 .
\end{equation}
Introducing Rindler coordinates $(t,\xi)$ through the transformations

\begin{equation}
T = \xi \sinh (a t), \qquad
X = \xi \cosh (a t),
\end{equation}
with $\xi > 0$ and constant proper acceleration parameter $a$, the metric takes the form

\begin{equation}
ds^2 = - (a \xi)^2 dt^2 + d\xi^2 + dY^2 + dZ^2 .
\end{equation}
These coordinates cover the right Rindler wedge $X>|T|$ and describe a congruence of observers undergoing uniform proper acceleration \cite{Rindler1}.

To understand the kinematics of accelerated observers, note that observers at fixed spatial coordinates $(\xi,Y,Z)$ follow worldlines with four-velocity

\begin{equation}
u^{\mu} = \frac{1}{a\xi}\,\delta^{\mu}_{\,t}.
\end{equation}
Their four-acceleration is given by (\ref{4-a}), with magnitude

\begin{equation}
|a| = \sqrt{a_{\mu} a^{\mu}} = \frac{1}{\xi}.
\end{equation}
Thus, each observer experiences a constant proper acceleration inversely proportional to its Rindler position $\xi$. In the Rindler description, the acceleration refers to the proper acceleration of observers with respect to inertial Minkowski frames. Although the spacetime remains flat, this acceleration induces an effective gravitational field in the non-inertial frame.

\begin{figure}[h]
\centering
\includegraphics[width=0.60\textwidth]{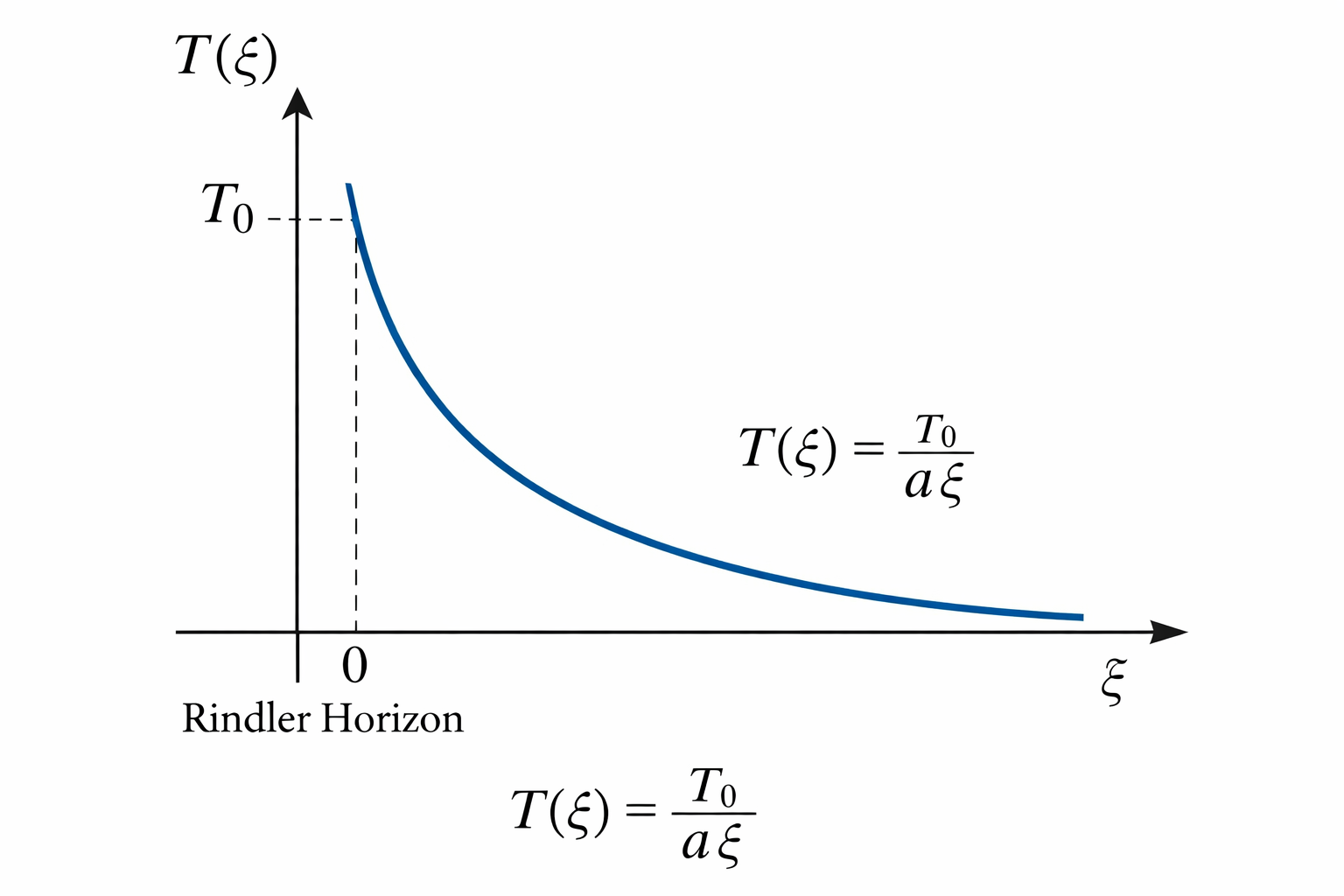}
\caption{The local thermodynamic temperature of a fluid at rest with respect to the Rindler frame.}
\label{fig1}
\end{figure}

The Rindler metric is stationary with $g_{00} = - (a\xi)^2$, and thus, by applying the Tolman--Ehrenfest relation $T(\xi)\, \sqrt{-g_{00}} = \text{const}$, one immediately finds

\begin{equation}
T(\xi) = \frac{T_0}{a \xi},
\end{equation}
where $T_0$ is a constant characterizing the equilibrium state. This constant is not a locally measured temperature, nor does it correspond to the temperature in an inertial frame. Rather, $T_0$ represents the Tolman constant associated with thermal equilibrium along the timelike Killing vector of the Rindler spacetime. It ensures the vanishing of heat flow in the accelerated frame and serves as a global parameter characterizing thermal equilibrium. This result shows that, in a uniformly accelerated frame, thermal equilibrium requires a spatially varying temperature profile that compensates the inertial effects associated with acceleration. The temperature gradient implied by the above relation satisfies

\begin{equation}
\nabla_{\xi} \ln T = - \frac{1}{\xi} = - |a| ,
\end{equation}
in agreement with the general equilibrium condition (\ref{A}). Therefore, the Rindler spacetime provides an explicit example in which acceleration plays the role of an effective gravitational field and thermal equilibrium is maintained through a position-dependent temperature. The temperature $T(\xi)$ represents the local thermodynamic temperature of a fluid that is at rest with respect to the Rindler frame. Each fluid element follows a uniformly accelerated worldline in Minkowski spacetime, while remaining static in Rindler coordinates. The temperature is measured by a comoving accelerated observer and varies with $\xi$ in such a way that thermal equilibrium is maintained in the accelerated frame. Note however that the Rindler frame possesses a causal horizon at $\xi=0$, beyond which the coordinates break down, see figure \ref{fig1}. 

While the Tolman law determines the thermodynamic temperature of matter in local equilibrium, accelerated observers also associate a temperature to the Minkowski vacuum, known as the Unruh temperature \cite{Unru1}

\begin{equation}
T_U = \frac{a}{2\pi}.
\end{equation}
This temperature has a fundamentally different origin and should not be confused with the equilibrium temperature $T(\xi)$ of the fluid.

In the following section, we investigate how these results are modified when the accelerated frame is filled with a relativistic fluid, with particular emphasis on the role of pressure and negative-pressure equations of state.

\section{Perfect fluids in uniformly accelerated frames}

In this section, we study the thermodynamic equilibrium of a relativistic perfect fluid as described by uniformly accelerated observers. Our goal is to determine how the standard conditions of equilibrium are modified in a non-inertial frame and to clarify the role played by pressure in the presence of acceleration. This section provides a dynamical realization of the thermal equilibrium discussed in previous section by explicitly modeling the system as a perfect fluid at rest in the Rindler frame.

To begin, we consider a relativistic perfect fluid characterized by the energy--momentum tensor

\begin{equation}
T^{\mu\nu} = (\rho + p)\, u^{\mu} u^{\nu} + p\, g^{\mu\nu},
\end{equation}
where $\rho$ is the energy density, $p$ the isotropic pressure and $u^{\mu}$ the four-velocity of the fluid elements.
In the Rindler frame introduced in the previous section, we assume that the fluid is at rest with respect to the accelerated observers. The four-velocity field is therefore aligned with the timelike Killing vector,
\begin{equation}
u^{\mu} = \frac{1}{\sqrt{-g_{00}}}\, \delta^{\mu}_{\,t}
= \frac{1}{a\xi}\, \delta^{\mu}_{\,t}.
\end{equation}
This choice corresponds to a stationary configuration with no macroscopic flow. Thermodynamic equilibrium requires the local conservation of energy and momentum

\begin{equation}
\nabla_{\mu} T^{\mu\nu} = 0.
\end{equation}
For a stationary fluid configuration, the nontrivial content of this equation resides in the spatial components $\nu=i$, which lead to the relativistic generalization of hydrostatic equilibrium.
A straightforward computation yields

\begin{equation}
\nabla_i p = - (\rho + p)\, a_i ,
\end{equation}
where $a_i = u^{\mu} \nabla_{\mu} u_i$ represents spatial components of the four-acceleration of the fluid elements in the comoving (Rindler) frame obtained by projection orthogonal to $u^{\mu}$.
In the Rindler frame, the only nonvanishing component of the acceleration is

\begin{equation}
a_{\xi} = \partial_{\xi} \ln \sqrt{-g_{00}} = \frac{1}{\xi},
\end{equation}
and the equilibrium condition takes the explicit form

\begin{equation}
\frac{d p}{d \xi} = - \frac{\rho + p}{\xi}.
\label{hydrostatic}
\end{equation}
Equation~\eqref{hydrostatic} represents the condition of hydrostatic equilibrium in a uniformly accelerated frame and is entirely kinematical, relying only on local conservation laws.

Now, let us consider the thermodynamic relations and local temperature. The local first law of thermodynamics for a perfect fluid can be written as

\begin{equation}
d\rho = T\, ds + \mu\, dn ,
\end{equation}
where $s$ is the entropy density, $n$ the particle number density and $\mu$ the chemical potential. In equilibrium, the Gibbs--Duhem relation reads

\begin{equation}
dp = s\, dT + n\, d\mu .
\end{equation}
Combining the hydrostatic equilibrium condition~\eqref{hydrostatic} with the Gibbs--Duhem relation, one finds

\begin{equation}
s\, \nabla_i T + n\, \nabla_i \mu = - (\rho + p)\, a_i .
\end{equation}
Using the thermodynamic identity

\begin{equation}
\rho + p = Ts + \mu n ,
\end{equation}
the above equation may be rewritten as

\begin{equation}
s \left( \nabla_i T + T a_i \right)
+ n \left( \nabla_i \mu + \mu a_i \right) = 0 .
\end{equation}
Since entropy and particle number are independently conserved in equilibrium, this condition implies

\begin{equation}
\nabla_i \ln T = - a_i,
\qquad
\nabla_i \ln \mu = - a_i .
\end{equation}
Integrating these relations yields

\begin{equation}
T \sqrt{-g_{00}} = \text{const},
\qquad
\mu \sqrt{-g_{00}} = \text{const},
\end{equation}
which is precisely the Tolman--Ehrenfest law and its chemical-potential analogue. For the Rindler metric, where $g_{00}=-(a\xi)^2$, the above relations lead to

\begin{equation}
T(\xi) = \frac{T_0}{a\xi},
\qquad
\mu(\xi) = \frac{\mu_0}{a\xi}.
\end{equation}
These expressions confirm that the spatial temperature profile obtained previously is fully consistent with the dynamical equilibrium conditions of a relativistic perfect fluid.
Importantly, the derivation does not rely on a specific equation of state and remains valid for arbitrary relations between $\rho$ and $p$.

On the role of pressure, note that the pressure enters the equilibrium condition only through the combination $\rho + p$, which controls the coupling between the fluid and the acceleration field. This observation will play a central role in the following section, where we examine fluids with negative pressure and discuss the physical interpretation of thermal equilibrium in such cases.
At this stage, we emphasize that the Tolman--Ehrenfest law remains formally valid for all perfect fluids, regardless of the sign of the pressure. However, the thermodynamic meaning and stability of the resulting equilibrium configurations require a more careful analysis.

\section{Fluids with negative pressure in accelerated frames}

We now turn to the case of relativistic fluids with negative pressure, which play a central role in cosmology and gravitational physics. Our aim is to examine how the equilibrium conditions derived in the previous section are modified when $p<0$, and to clarify the physical meaning of temperature in such systems as perceived by accelerated observers.

For a perfect fluid at rest in the Rindler frame, the condition of hydrostatic equilibrium was shown to take the form of Eq. (\ref{hydrostatic}), which remains formally valid for all perfect fluids, irrespective of the sign of the pressure. However, for fluids with negative pressure, the combination $\rho + p$ may become small, vanish, or even change sign. Since this combination controls the coupling between the fluid and the acceleration field, its sign has important physical consequences.

${\bullet}$ {\it{Case $\rho + p > 0$}}: For ordinary matter and radiation, as well as for many exotic fluids with moderately negative pressure, one has $\rho + p > 0$. In this case, Eq.~\eqref{hydrostatic} implies that the pressure decreases monotonically with increasing $\xi$, in direct analogy with standard hydrostatic equilibrium in a gravitational field.
The Tolman--Ehrenfest law $T(\xi) = \frac{T_0}{a\xi}$, is compatible with a stable equilibrium configuration, and the temperature gradient compensates the inertial force associated with acceleration.

${\bullet}$ {\it{Case $\rho + p=0, vacuum- like fluids$}}: A particularly important case is that of vacuum-like fluids satisfying

\begin{equation}
p = - \rho .
\end{equation}
This equation of state characterizes a cosmological constant or de Sitter–like fluid. In this case, Eq.~\eqref{hydrostatic} reduces to

\begin{equation}
\frac{d p}{d \xi} = 0,
\end{equation}
implying that both $\rho$ and $p$ are spatially constant in the accelerated frame. The fluid does not respond to the acceleration field, reflecting the absence of inertial mass density.
Although the Tolman law formally yields a position-dependent temperature $T(\xi) = \frac{T_0}{a\xi}$, its thermodynamic interpretation becomes subtle. In particular, since $\rho + p = 0$, the Gibbs relation implies

\begin{equation}
Ts + \mu n = 0,
\end{equation}
which severely restricts the allowed thermodynamic degrees of freedom. In this sense, the notion of temperature loses its conventional meaning for vacuum-like fluids.

${\bullet}$ {\it{Case $\rho + p<0, instability$}}: For fluids satisfying $\rho + p < 0$, the right-hand side of Eq.~\eqref{hydrostatic} changes sign, leading to

\begin{equation}
\frac{d p}{d \xi} > 0 .
\end{equation}
In this regime, pressure increases with $\xi$, indicating that acceleration acts as an effective repulsive force.
Such configurations are generically unstable against perturbations, as the equilibrium condition requires a temperature gradient that amplifies, rather than compensates, the inertial effects. Although the Tolman--Ehrenfest relation remains mathematically valid, the resulting equilibrium state cannot be interpreted as thermodynamically stable.
This observation highlights the fact that the Tolman law is a necessary but not sufficient condition for physical equilibrium in accelerated frames.

For fluids with negative pressure, the distinction between thermodynamic temperature and operational temperature becomes particularly important. While the Tolman law determines the former, an accelerated observer also perceives a temperature associated with vacuum fluctuations, namely the Unruh temperature $T_U = \frac{a}{2\pi}$.
For vacuum-like fluids, it is the Unruh temperature, rather than the fluid temperature, that governs the response of particle detectors. This reinforces the view that the temperature appearing in the Tolman relation should be regarded as a formal thermodynamic parameter, whose physical measurability depends on the nature of the underlying degrees of freedom.

The analysis above shows that, although the Tolman--Ehrenfest law holds universally for perfect fluids in accelerated frames, its physical interpretation crucially depends on the sign of $\rho + p$. For ordinary fluids, it describes a stable equilibrium configuration, whereas for vacuum-like or phantom fluids the concept of temperature becomes ambiguous or physically inaccessible.
These results provide a natural bridge between relativistic thermodynamics, accelerated observers, and cosmological fluids, and will be further discussed in the concluding section.

\section{Quantum perspective: Unruh effect and particle detectors}

So far, our analysis has been entirely classical and thermodynamic, focusing on the conditions for thermal equilibrium of a fluid at rest in an accelerated (Rindler) frame. Now, let us briefly discuss the quantum-field-theoretic perspective on accelerated observers, in particular the Unruh effect, and clarify its conceptual relation to the temperature profiles obtained in the previous sections.

A uniformly accelerated observer with proper acceleration $a$ in Minkowski spacetime perceives the Minkowski vacuum as a thermal state with the Unruh temperature \cite{Unru1, crispino2008}

\begin{equation}
T_U = \frac{\hbar a}{2\pi k_B}.
\end{equation}
This result is purely quantum mechanical and does not rely on the presence of any material medium or fluid. The thermal character emerges from the observer-dependent notion of particles and the restriction of field modes to the Rindler wedge. It is important to emphasize that $T_U$ does not represent the temperature of a physical fluid in thermal equilibrium, but rather an effective temperature characterizing the response of an accelerated detector coupled to a quantum field.

A convenient operational way to illustrate the Unruh effect is provided by the Unruh--DeWitt detector model \cite{unruh1976, louko2008, takagi1986}. The detector is idealized as a two-level quantum system with energy gap $\omega$, linearly coupled to a scalar field $\phi(x)$ along its worldline $x^\mu(\tau)$. The interaction Hamiltonian is given by \cite{dewitt1979}

\begin{equation}
H_{\text{int}} = \lambda\, m(\tau)\, \phi[x(\tau)],
\end{equation}
where $\lambda$ is a small coupling constant and $m(\tau)$ is the detector monopole operator.
To leading order in perturbation theory, the transition probability per unit proper time from the ground state to the excited state is proportional to the response function

\begin{equation}
\dot{P}(\omega) \propto \int_{-\infty}^{\infty} d\Delta\tau \,
e^{-i\omega \Delta\tau}\,
G^{+}\!\left(\Delta\tau\right),
\end{equation}
where $G^{+}$ is the Wightman function of the field evaluated along the detector trajectory. For a uniformly accelerated trajectory

\begin{equation}
t(\tau) = \frac{1}{a}\sinh(a\tau), \qquad
x(\tau) = \frac{1}{a}\cosh(a\tau),
\end{equation}
the response function takes a thermal form \cite{birrelldavies} 

\begin{equation}
\dot{P}(\omega) \propto
\frac{\omega}{e^{2\pi\omega/a}-1},
\end{equation}
corresponding to a Planckian spectrum at the Unruh temperature $T_U$.

The quantum temperature $T_U$ should be clearly distinguished from the classical temperature profile $T(\xi)$ obtained in the previous sections. The latter arises from the condition of local thermal equilibrium for a fluid at rest in the accelerated frame and satisfies $\nabla_i \ln T = - a_i$, leading, in the Rindler case, to a position-dependent temperature $T(\xi)$.
In contrast, the Unruh temperature characterizes the response of a quantum detector in the Minkowski vacuum and does not rely on any equation of state or thermodynamic relations. While both temperatures are associated with acceleration, they originate from fundamentally different physical mechanisms.

The comparison highlights that acceleration can give rise to thermal features both in classical relativistic thermodynamics and in quantum field theory, but with distinct interpretations. In the present work, the temperature profile is a manifestation of equilibrium conditions in a non-inertial frame, whereas the Unruh temperature reflects the observer-dependent particle content of quantum fields. These two notions of temperature should therefore not be identified in general.

\section{Conclusion}

In this work, we have investigated the notion of temperature and thermal equilibrium from the perspective of uniformly accelerated observers, focusing on the Rindler frame as a paradigmatic example of a non-inertial reference system. Our analysis has been carried out entirely within the framework of relativistic classical thermodynamics and relativistic fluid dynamics, without invoking specific microscopic models or equations of state.

Starting from the covariant conservation of the energy--momentum tensor, we derived the condition for hydrostatic equilibrium of a perfect fluid in an accelerated frame. We showed that the presence of acceleration necessarily leads to a spatially varying temperature profile, which is determined by the four-acceleration of the fluid worldlines. In the Rindler spacetime, this condition yields a temperature inversely proportional to the Rindler spatial coordinate, in direct analogy with the Tolman--Ehrenfest relation in static gravitational fields.

A key conceptual point of this work is that the resulting temperature field should not be interpreted as the outcome of a transformation law between different observers. Unlike the case of Lorentz transformations between inertial frames, where one compares temperatures assigned to the same fluid state by different observers, the present analysis concerns a single physical configuration: a fluid in local thermal equilibrium that is at rest with respect to an accelerated frame. The temperature profile emerges as a consistency condition for equilibrium in the presence of acceleration, rather than as an observer-dependent quantity obtained by transforming a global temperature.

This distinction highlights a fundamental difference between relativistic thermodynamics in inertial and non-inertial settings. In accelerated frames, equilibrium itself is intrinsically inhomogeneous, and temperature must be regarded as a local scalar field tied to the fluid four-velocity and its acceleration. Within this framework, no universal observer-to-observer transformation law for temperature is required or even meaningful.

It is worth emphasizing that the distinction between acceleration-induced effects and gravitational effects should be understood with some care. In general relativity, inertial and gravitational effects cannot always be separated globally in an invariant manner, as illustrated by the equivalence principle and by the appearance of inertial forces in accelerated or rotating reference frames. Coriolis and centrifugal effects in rotating frames provide familiar examples of this broader issue. In the present work, however, we do not attempt to make a global distinction between `inertial'' and `gravitational'' thermal effects. Our analysis is restricted to a uniformly accelerated frame in flat spacetime and concerns the local thermodynamic equilibrium of a perfect fluid with respect to that frame. The temperature profile derived above is therefore a statement about the equilibrium state associated with the specified congruence of accelerated observers, independently of whether the corresponding inertial effects are interpreted in an alternative gravitational language.

Although our treatment is classical, the results naturally invite comparison with quantum effects associated with acceleration, most notably the Unruh effect. While the Unruh temperature characterizes the response of quantum detectors in accelerated motion through the Minkowski vacuum, the temperature field derived here describes the thermodynamic equilibrium of classical matter in an accelerated frame. The formal similarities between these two notions of temperature underscore the deep interplay between acceleration, thermodynamics and quantum field theory, while also emphasizing their distinct physical origins.

The present work may be extended in several directions. Possible developments include the study of dissipative fluids in accelerated frames, the role of entropy production and transport phenomena, and the extension to genuinely curved spacetimes beyond the Rindler approximation. A more detailed comparison between classical equilibrium temperature profiles and quantum detector-based temperatures may also provide further insight into the thermodynamics of accelerated systems.

\end{document}